\documentclass[11pt,a4paper]{article}

\usepackage[utf8]{inputenc}
\usepackage[T1]{fontenc}
\IfFileExists{lmodern.sty}{\usepackage{lmodern}}{\usepackage{times}}
\usepackage[margin=1in]{geometry}
\usepackage{amsmath,amssymb}
\usepackage{graphicx}
\usepackage{siunitx}
\usepackage{authblk}
\usepackage[numbers,sort&compress]{natbib}
\usepackage{caption}
\usepackage{xcolor}
\usepackage[colorlinks=true,allcolors=blue]{hyperref}
\usepackage{microtype}

\DeclareSIUnit{\torr}{Torr}
\DeclareSIUnit{\angstrom}{\text{\AA}}

\title{\bfseries Low Loss Superconducting Resonators Enabled by Aluminum
Microstructural Engineering and Dielectric Trimming}

\author[1,2]{Mahmoud Almansouri\thanks{Corresponding author: \href{mailto:mahmoud.almansouri@kaust.edu.sa}{mahmoud.almansouri@kaust.edu.sa}}}
\author[2]{Umar T. F. Alhuwaymel}
\author[2]{Ibraheem AlYousef}
\author[2]{Albaraa Shafi}
\author[2]{Abdullah Albogime}
\author[1]{Igor Getmanov}
\author[1]{Ahmed Hajr}
\author[1]{Atif Shamim}

\affil[1]{Physical Science and Engineering, King Abdullah University of Science and Technology, Thuwal, Saudi Arabia}
\affil[2]{Research and Development, National Company of Telecommunications and Information Security, Jeddah, Saudi Arabia}

\date{}

\begin{document}

\maketitle

\begin{abstract}
\noindent
Material losses in superconducting circuits fundamentally limit qubit coherence
times and resonator quality factors. Most research efforts focus on mitigating
losses at circuit interfaces, including metal--substrate, substrate--air, and
metal--air interfaces. However, the correlation between TLS and non-TLS losses
with the intrinsic properties of the superconducting metal and the dielectric
edge smoothness is not well studied. In this work, we link the aluminum film
grain size to non-TLS losses and the dielectric trimming profile and roughness
to TLS loss; both loss mechanisms are subsequently mitigated. To reduce
metal-related losses, we engineer the aluminum microstructure by heating during
deposition, increasing grain size and reducing grain boundary density. Beyond
mitigating metal losses, we introduce a two-step etching technique, Tropic
etching, to suppress dielectric TLS loss by producing an ultra-smooth silicon
surface with minimal defects and redeposition. These results lay out the
fabrication pathway for aluminum resonators with lower loss, demonstrating
two-orders-of-magnitude improvement in quality factor from $6\times10^{4}$ to
$2.3\times10^{6}$. Since aluminum is the basis for most high-coherence
Josephson junctions and dielectric edges are inherent to all common device
geometries, the improvements in aluminum microstructure and edge profiling,
presented here, can enhance the performance of superconducting quantum devices.
\end{abstract}


\vspace{1em}

\section{Introduction}

Superconducting quantum circuits are a leading platform for quantum
computation, enabled by fast gate operations and high-fidelity control and
measurements \cite{barends2014,kim2023,kjaergaard2020,acharya2025,houck2008,devoret2004}.
Their compatibility with established semiconductor fabrication processes and
microwave readout systems further facilitates scalable integration
\cite{bardin2021,verjauw2022,brecht2016}. Superconducting quantum computing has
come a long way since its initial demonstrations, owing to advances in circuit
design, nanofabrication, and materials engineering that have dramatically
improved qubit coherence and control \cite{wang2022,place2021}. These
developments have extended the timescales over which quantum information can be
stored and manipulated, enabling milestone demonstrations on the path toward
quantum advantage \cite{kim2023,arute2019,google2025,sivak2023}.

Superconducting qubits coupled to microwave resonators form the fundamental
building blocks of these circuits, enabling quantum information processing and
qubit readout \cite{blais2021,wallraff2004}. The performance of both components
is limited by energy dissipation at cryogenic temperatures. In particular, qubit
coherence times and resonator quality factors are strongly affected by microwave
losses arising from material and fabrication imperfections
\cite{martinis2005,siddiqi2021,ganjam2024}. A dominant contribution at cryogenic
temperatures comes from two-level-system (TLS) defects---forms of material
disorder such as dangling bonds, vacancies, and charged dipoles that reside in
amorphous oxides and interfaces \cite{phillips1972,anderson1972,muller2019,gao2008}.
In contrast, non-TLS losses are typically tied to the superconductor, such as
quasiparticle generation, magnetic flux vortices, and Cooper-pair breaking by
stray photons or substrate phonons
\cite{catelani2011,song2009,chiaro2016,liu2024}. Together, these mechanisms
bound the performance of present-day superconducting quantum circuits
\cite{siddiqi2021,oliver2013}.

Superconducting resonators are used in quantum circuits to read out qubit
states, inherently filtering microwave noise through their narrow bandwidth and
high quality factor \cite{krantz2019,deleon2021}. They also serve as sensitive
probes for characterizing TLS and non-TLS losses through quality factor
measurements \cite{gao2008,lei2023,mcrae2020}, providing a convenient testbed
for detecting and mitigating both contributions. Non-TLS losses have been
suppressed through approaches such as flux-trapping geometries and improved
magnetic shielding \cite{chiaro2016,barends2011}, while TLS contributions have
been reduced through surface engineering and material substitution
\cite{cwang2015,calusine2018,place2021}. Despite this progress, the individual
contributions of specific resonator physical parameters---in particular
superconductor microstructure and dielectric surface roughness---to the overall
quality factor remain poorly understood. In this work, we focus on these two
parameters to improve aluminum-resonator quality factors.

Aluminum thin films are among the most widely used materials in superconducting
quantum circuits, serving as the ground plane and qubit capacitor pad
metallization, and offering straightforward integration with Josephson junctions
\cite{zeng2016,chayanun2024,burnett2019,carroll2022}. Understanding how aluminum
microstructure---particularly grain size---affects loss is therefore essential
for improving circuit performance \cite{mcrae2020,murray2021,biznarova2024}.
Previous efforts have explored the columnar growth of aluminum thin film grains
by increasing the film thickness from 150 to \SI{500}{\nano\meter} to reduce the
density of grain boundaries, improving the resonator quality factor from
$9\times10^{5}$ to $1.4\times10^{6}$ at \SI{4.45}{\giga\hertz}
\cite{biznarova2024}. In contrast, a recent investigation shows that depositing
aluminum at a cryogenic temperature of \SI{6}{\kelvin}---which reduces the grain
size and increases the total number of grain boundaries---yields no measurable
improvement in quality factor compared to room-temperature deposition
\cite{aguilar2026}. Together, these investigations have reported either an
improvement in quality factor at a single frequency or no clear correlation
between aluminum microstructure and resonator quality factor, leaving the
underlying relationship poorly understood. Here, we address this gap by heating
the substrate during aluminum deposition to reduce the grain boundary density,
and we study the resulting improvement in average quality factor across
resonators spanning a \SI{1.5}{\giga\hertz} frequency range.

In addition to metallic losses, dielectric losses at material interfaces play a
central role in limiting the performance of superconducting resonators, with
defects, native oxides, and fabrication byproducts at the metal--air (MA),
substrate--air (SA), and metal--substrate (MS) interfaces identified as dominant
loss contributors \cite{murray2021,altoe2022,wenner2011}. Significant efforts
have therefore focused on mitigating these losses through interface engineering,
materials optimization, and deep etching strategies aimed at reducing the
electromagnetic field participation in lossy dielectric regions
\cite{calusine2018,woods2019,chu2016}. In coplanar waveguide (CPW) resonators,
the center conductor and ground planes lie on the same substrate surface, and
the electromagnetic field is strongly concentrated at the metal edges where the
center conductor meets the substrate. The SA interface adjacent to these edges
is therefore a particularly significant source of dielectric loss
\cite{wenner2011}. Among the most effective fabrication strategies for
mitigating this contribution is isotropic deep silicon etching beneath and
beside the metal edges, which physically removes lossy substrate material from
the high-field region and thereby lowers the SA participation ratio
\cite{zikiy2023}. Isotropic silicon etching for this purpose is performed almost
exclusively with SF$_6$-based reactive ion etching (RIE), as it is one of the few
processes that delivers the lateral undercut required by the trenched CPW
geometry. However, SF$_6$ etching leaves sidewall and bottom roughness arising
from silicon redeposition during the etch process \cite{muller2019,larsen2006}.
This roughness acts as TLS defect sites and reintroduces dielectric
participation in the trenched region. Trenched CPW geometries therefore reduce
SA participation but reintroduce loss through etch-induced surface roughness.
Mitigating this roughness from SF$_6$-based isotropic etching offers a route to
further improving the internal quality factor of trenched superconducting
resonators---a direction we pursue in this work.

We address the two gaps identified above, aluminum grain microstructure and
silicon dielectric smoothness, by examining their impact on the resonator
quality factor and isolating their respective contributions to non-TLS and TLS
loss. We mitigate both by reducing the aluminum grain boundary density and
applying a CHF$_3$ follow-up etch that smooths the dielectric surface while
preserving the SF$_6$ undercut profile. To suppress non-TLS loss, we use
substrate heating during aluminum deposition to reduce the grain boundary
density and increase the average grain size to \SI{1.2}{\micro\meter}. To
suppress TLS loss, we introduce a two-step etching technique, Tropic etch, in
which an SF$_6$ isotropic etch is followed by a CHF$_3$ step that smooths the
trenched surfaces while preserving the undercut geometry. By combining these
loss-mitigation strategies, we demonstrate a 40-fold improvement in the internal
quality factor of aluminum resonators, from $6\times10^{4}$ in untreated devices
to $2.3\times10^{6}$ with both techniques applied. These results establish
grain-boundary reduction in the superconductor and dielectric trimming as
complementary strategies for reducing loss in superconducting resonators. Both
strategies are expected to generalize beyond the aluminum-on-silicon platform
demonstrated here, offering a broadly applicable route to reducing loss in
superconducting quantum circuits.

\section{Results and Discussion}

\begin{figure}[!ht]
\centering
\includegraphics[width=\textwidth]{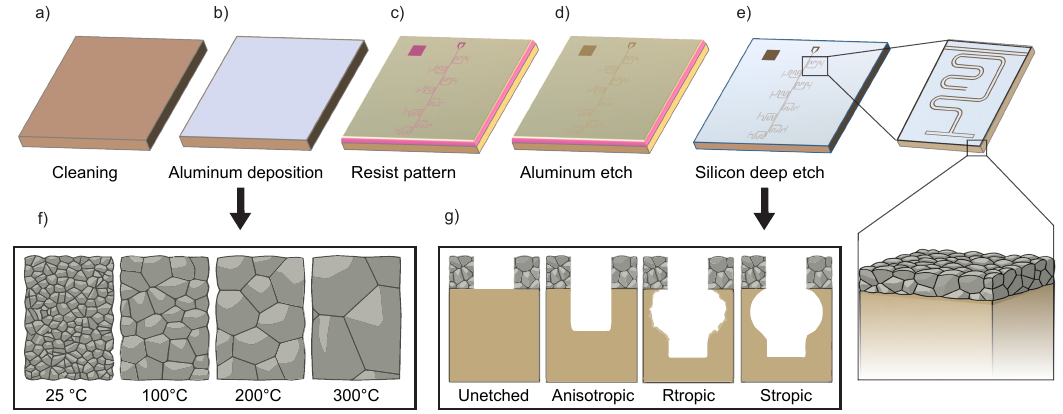}
\caption{Fabrication process for the resonators, combining microstructural
engineering and dielectric trimming. (a) A high-resistivity silicon wafer is
cleaned and stripped of native oxide [using buffered oxide etch (BOE 7:1) / in
dilute HF]. (b) High-purity aluminum is deposited at different substrate
temperatures to engineer the film microstructure, with the inset (f) showing the
conditions used: RT, \SI{100}{\celsius}, \SI{200}{\celsius}, and
\SI{300}{\celsius}. (c) The circuitry is defined by photolithography. (d) The
aluminum is dry-etched using a Cl$_2$/BCl$_3$ reactive-ion etch (RIE). (e) The
exposed silicon is deep-etched by RIE using CHF$_3$ or a combination of SF$_6$
and CHF$_3$, producing the different edge profiles shown in inset (g).}
\label{fig:fab}
\end{figure}

We implement aluminum microstructural engineering and silicon dielectric
trimming to fabricate low-loss aluminum coplanar waveguide resonators on
high-resistivity silicon. Figure~\ref{fig:fab} illustrates the full fabrication
sequence. All resonators share the same process flow, with aluminum deposited on
each of four wafers at a different substrate temperature. Within each wafer,
resonators are deep etched under different dielectric trimming conditions to
vary the silicon edge profile and surface roughness. The metal--air (MA) and
metal--substrate (MS) interfaces are therefore the same across all wafers, while
the substrate--air (SA) interface differs depending on the dielectric trimming
applied. The aluminum microstructural engineering is implemented in the
deposition step after wafer cleaning (Figure~\ref{fig:fab}b), progressively
increasing the grain size and reducing the grain boundary density. The
dielectric trimming, on the other hand, is performed at the silicon deep-etching
step (Figure~\ref{fig:fab}e), after aluminum etching (Figure~\ref{fig:fab}d). We
first examine the role of the aluminum microstructure in this subsection, then
turn to dielectric trimming of the substrate--air interface.

\subsection{Aluminum Microstructural Engineering}

\begin{figure}[!ht]
\centering
\includegraphics[width=0.75\textwidth]{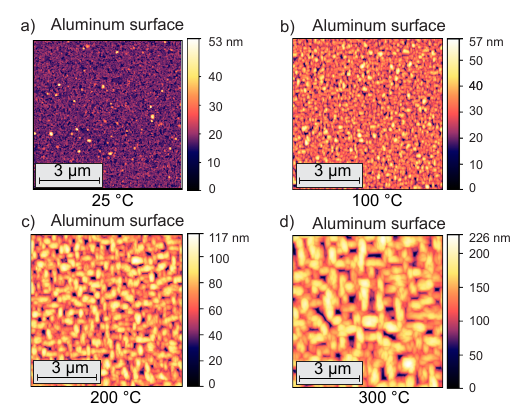}
\caption{Microstructural engineering of the aluminum grains. Atomic force
microscopy (AFM) images showing the evolution of the aluminum grain morphology
with increasing substrate temperature: (a) room temperature (RT), (b)
\SI{100}{\celsius}, (c) \SI{200}{\celsius}, and (d) \SI{300}{\celsius}.}
\label{fig:afm}
\end{figure}

\begin{figure}[!ht]
\centering
\includegraphics[width=0.6\textwidth]{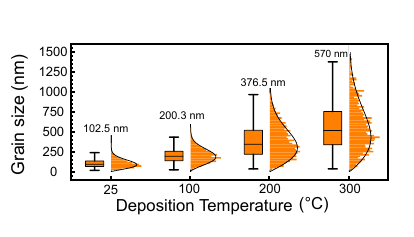}
\caption{Grain size distribution for the four deposition temperatures; the value
above each distribution gives the average grain size.}
\label{fig:grain}
\end{figure}

We deposit four \SI{200}{\nano\meter} aluminum films at substrate temperatures of
room temperature (RT), \SI{100}{\celsius}, \SI{200}{\celsius}, and
\SI{300}{\celsius}, and characterize the resulting microstructure.
Figure~\ref{fig:afm}(a--d) shows atomic force microscopy (AFM) images of the four
films, in which the aluminum grain size increases progressively with deposition
temperature, leaving fewer grain boundaries within the film. The average grain
size, extracted from the AFM images, increases from approximately
\SI{100}{\nano\meter} at room temperature to 200, 380, and
\SI{570}{\nano\meter} at 100, 200, and \SI{300}{\celsius}, respectively
(Figure~\ref{fig:grain}).

\begin{figure}[!ht]
\centering
\includegraphics[width=\textwidth]{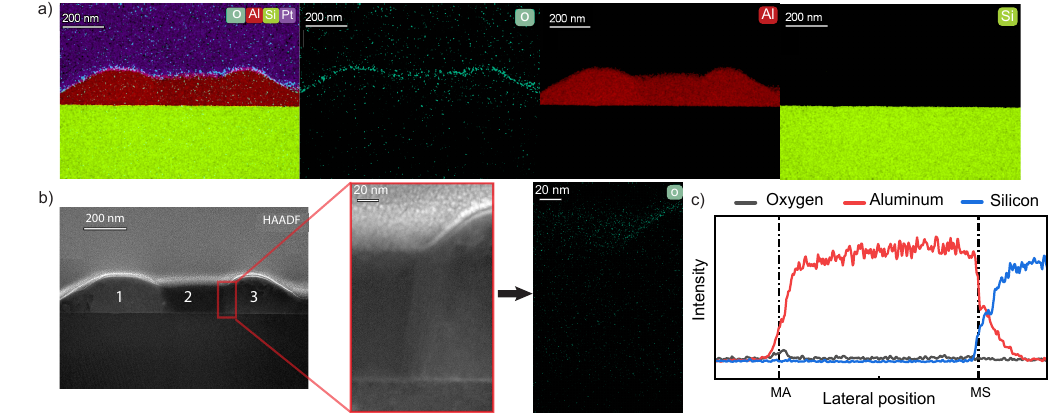}
\caption{Cross-sectional STEM analysis of the \SI{300}{\celsius} large-grain
aluminum film on silicon. (a) False-colored energy-dispersive X-ray spectroscopy
(EDS) maps showing the Al, O, Si, and Pt distributions across the sample
cross-section; the Pt layer is deposited during sample preparation. (b)
High-angle annular dark-field (HAADF) image of three aluminum grains (labeled 1,
2, 3), with the inset showing a zoomed-in HAADF image at the grain boundary
overlaid with the oxygen EDS map. (c) Elemental distribution along the
cross-section, with vertical lines marking the MA and MS interfaces.}
\label{fig:stem}
\end{figure}

Energy-dispersive X-ray spectroscopy (EDS) confirms that the SM interface is free
of oxide, with oxidation confined to the top of the aluminum film at the
metal--air (MA) interface. Oxides along aluminum grain boundaries have been
identified as a source of loss in the aluminum microstructure
\cite{biznarova2024}. To test for this directly, we examined the boundaries
between large grains in cross-section. Figure~\ref{fig:stem}(a) shows a
high-angle annular dark-field (HAADF) image of three large aluminum grains
(1, 2, 3); the boundary between grains 2 and 3 is magnified in the inset of
Figure~\ref{fig:stem}(b). EDS mapping across this boundary, at a spatial
resolution of \SI{20}{\nano\meter}, reveals no detectable oxide. The
compositional line profile in Figure~\ref{fig:stem}(c) corroborates this: the SM
interface and the film bulk are oxide-free, and an oxide signal appears only at
the MA interface. These observations enhance clarity on the microscopic origin of
losses in grain boundaries. TLS loss is predominantly associated with amorphous
oxides, particularly silicon oxides, whereas non-TLS loss correlates with the
superconducting film \cite{altoe2022}. Therefore, the absence of oxide at the
grain boundaries indicates that grain-boundary loss is non-TLS in nature. We
establish this in the discussion section through quality-factor analysis.

\subsection{Dielectric Trimming and Silicon Edge Profile}

\begin{figure}[!ht]
\centering
\includegraphics[width=0.8\textwidth]{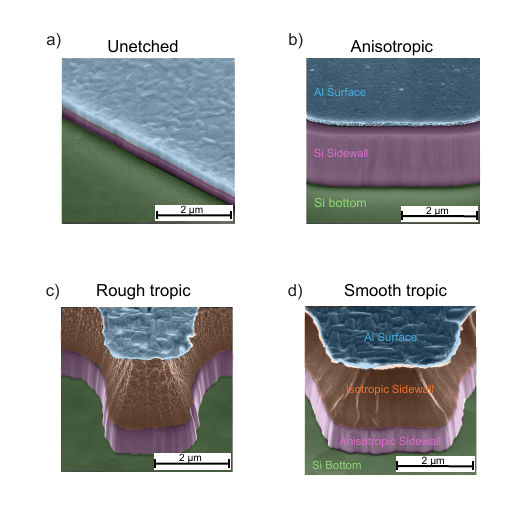}
\caption{Dielectric trimming of the coplanar waveguide edge. False-colored
scanning electron microscopy (SEM) images of the silicon edge profile for edges
fabricated with (a) aluminum etching only, (b) anisotropic deep etching with
CHF$_3$ following the aluminum etch, (c) the rough-sidewall Tropic etch (with
\SI{60}{\milli\torr} pressure during SF$_6$ etch followed by CHF$_3$;
`Rtropic'), and (d) the smooth-sidewall Tropic etch (with \SI{5}{\milli\torr}
pressure during SF$_6$ etch followed by CHF$_3$; `Stropic').}
\label{fig:edge}
\end{figure}

As discussed in the introduction, conventional SF$_6$ etching inherently roughens
the silicon surface, introducing surface contamination and TLS loss. To reduce SA
losses, we introduce dielectric trimming to remove the bottom roughness left by
deep etching while preserving the isotropic edge profile.
Figure~\ref{fig:edge}(a) shows false-colored scanning electron microscopy (SEM)
images of the baseline (Unetched) condition, in which only the aluminum is etched
after patterning by standard Cl$_2$/BCl$_3$ reactive-ion etching (RIE), with no
deep silicon etch applied. This step nonetheless removes a small amount of
silicon, visible as the purple region. The anisotropic CHF$_3$ etch in
Figure~\ref{fig:edge}(b) produces smooth, near-vertical sidewalls and a smooth
bottom surface. To further reduce SA loss, the Tropic etch combines the isotropic
profile with the CHF$_3$ smoothing step, yielding a smooth bottom surface
together with isotropic sidewalls. We apply it in two variants---a rough
(Rtropic) and a smooth (Stropic) condition, shown in Figure~\ref{fig:edge}(c,d)
---which share the same profile and differ only in sidewall roughness, set by the
RIE chamber pressure (\SI{60}{\milli\torr} versus \SI{5}{\milli\torr}; see
Supplementary Note~S1).

\begin{figure}[!ht]
\centering
\includegraphics[width=0.7\textwidth]{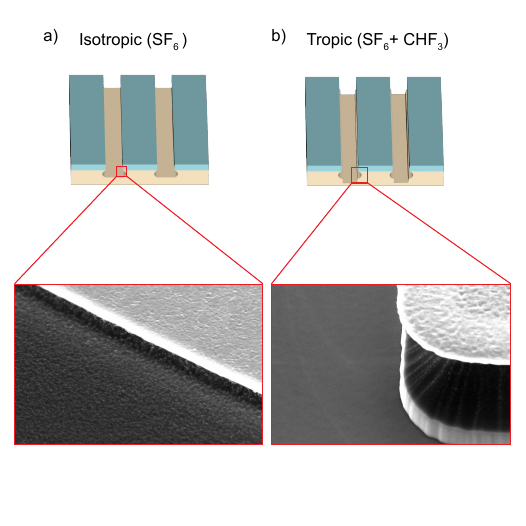}
\caption{Bottom-surface roughness of the silicon for the two etch profiles.
Schematic of the etched silicon profile with insets showing scanning electron
microscopy (SEM) images of the bottom surface. (a) Isotropic SF$_6$ etch, with
the inset showing the resulting bottom-surface roughness. (b) Tropic etch, with
the inset showing the smooth bottom surface obtained after the subsequent
CHF$_3$ step.}
\label{fig:bottom}
\end{figure}

The effectiveness of the smoothing step is shown in Figure~\ref{fig:bottom},
which compares the silicon bottom surface after an SF$_6$-only etch with that of
the Tropic etch. As seen in the SEM images, the subsequent CHF$_3$ step
substantially reduces the bottom-surface roughness, confirming that the hybrid
process recovers a smooth bottom surface while retaining the isotropic profile.

\begin{figure}[!ht]
\centering
\includegraphics[width=0.6\textwidth]{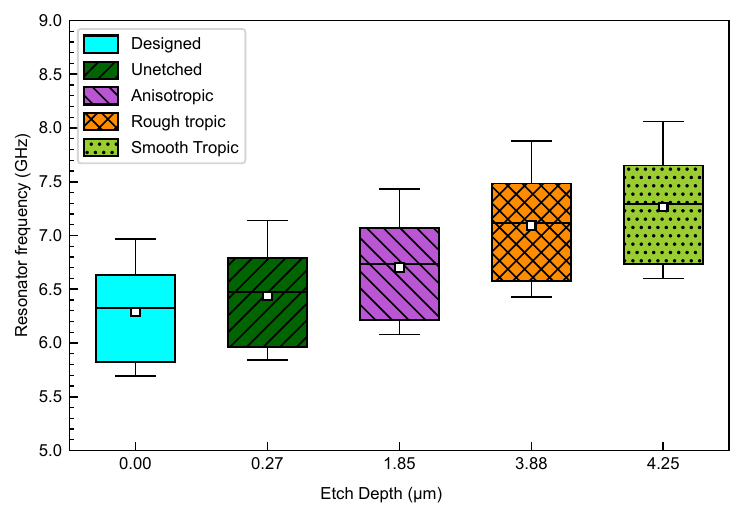}
\caption{Measured resonator frequency versus etch depth for the four etch
profiles. Resonator frequency of the $\lambda/4$ CPW resonators is plotted
against etch depth, measured by profilometry, for the unetched, anisotropic,
Rtropic, and Stropic conditions (legend). Each box summarizes the eight
resonators on a device: the box spans the frequency range, the line marks the
median, and the whiskers extend to the full range.}
\label{fig:freq}
\end{figure}

The edge profiles shown in Figure~\ref{fig:edge} were produced using different
etching parameters, yielding different etch depths. Removing dielectric from the
gaps lowers the effective permittivity $\epsilon_{\mathrm{eff}}$ of the coplanar
waveguide, and the resonant frequency therefore increases with etch depth. This
dependence is presented in Figure~\ref{fig:freq}. Since the frequency shift is a
function of etch depth, it can be exploited as a post-fabrication tuning
mechanism. By masking the metal with a protective resist and locally deep-etching
the exposed substrate, individual resonators can be trimmed toward their target
frequencies after the base layer has been defined. This provides a practical
means of adjusting individual resonator frequencies while leaving the resonator
design and total length unchanged.

\subsection{Superconducting Resonator Quality Factor}

\begin{figure}[!ht]
\centering
\includegraphics[width=0.75\textwidth]{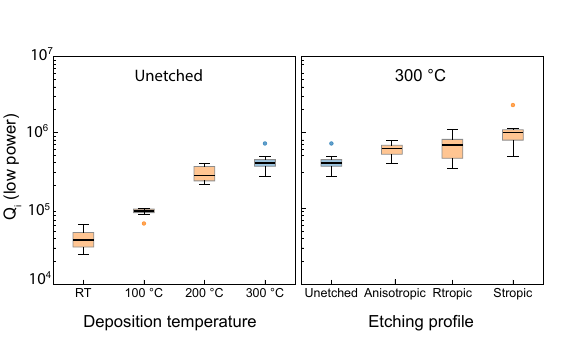}
\caption{Single-photon internal quality factor $Q_i$ of the CPW resonators. Left
panel: low-power $Q_i$ as a function of aluminum deposition temperature. Right
panel: low-power $Q_i$ of the \SI{300}{\celsius} film for the four etch profiles.
Each box summarizes the measured quality factors of at least eight resonators on
a device. The unetched \SI{300}{\celsius} device (blue) appears in both panels
and represents the same set of measurements.}
\label{fig:qi}
\end{figure}

To quantify the effectiveness of the proposed fabrication techniques, we
patterned resonators on the four wafers with different etch profiles and measured
their internal quality factors $Q_i$ at cryogenic temperatures (see Supplementary
Note~S2 for full measured data). All values reported in Figure~\ref{fig:qi} are
in the single-photon regime relevant to qubit operation. The left panel shows
$Q_i$ as a function of aluminum deposition temperature: the mean $Q_i$ increases
from $4\times10^{4}$ at room temperature to $4.2\times10^{5}$ at
\SI{300}{\celsius}, consistent with the reduced grain-boundary density of the
larger-grain films. We then applied the dielectric-trimming profiles to the
highest-performing \SI{300}{\celsius} film to reduce SA-interface loss. The right
panel shows the resulting progression, from the unetched \SI{300}{\celsius}
baseline to $2.3\times10^{6}$ for the Stropic condition---an overall improvement
of a factor of 40 times over the unetched baseline through the two fabrication
techniques introduced in this work.

These results indicate that silicon sidewall and bottom-surface roughness also
contribute to dielectric loss. Because the etch depths of Rtropic and Stropic are
comparable (3.9 versus \SI{4.2}{\micro\meter}), we attribute the difference in
$Q_i$ between them to sidewall smoothing: Stropic roughly doubles the $Q_i$ of
Rtropic through surface smoothing alone. To rule out etch depth as the source of
the improvement, we additionally measured an Stropic device etched to
\SI{3}{\micro\meter} rather than \SI{4.2}{\micro\meter} and found consistent
performance (see Supplementary Note~S3). Together, these comparisons indicate
that the $Q_i$ improvement arises from larger grains and the edge profile and
silicon smoothness introduced here, rather than from etch depth.

\subsection{Loss Decomposition and Mechanisms}

\begin{figure}[!ht]
\centering
\includegraphics[width=\textwidth]{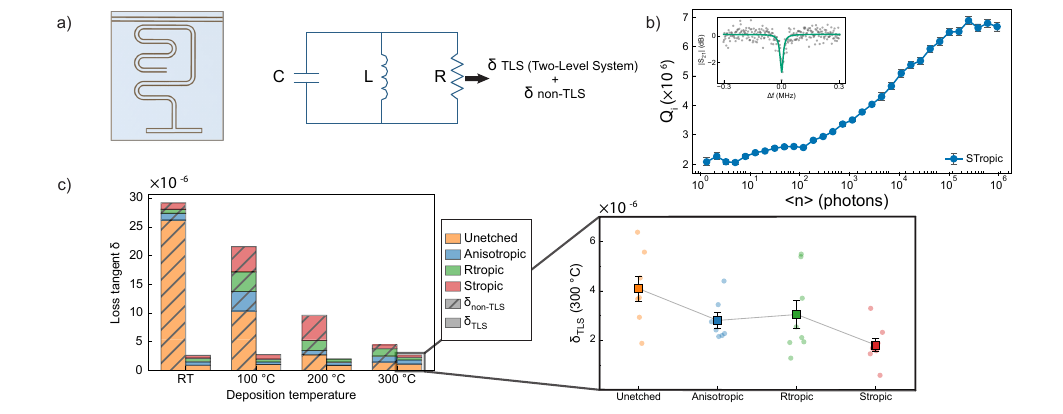}
\caption{Reduction of the TLS and non-TLS loss in the resonators. (a) Schematic
of the CPW resonator and its equivalent RLC circuit model, in which the
resistance $R$ represents the total internal loss, decomposed into
two-level-system (TLS) and non-TLS contributions. (b) Power dependence of the
internal quality factor $Q_i$ for the best-performing resonator
(\SI{300}{\celsius} film, Stropic etch), reaching a single-photon $Q_i$ of
$2.3\times10^{6}$; the inset shows a representative single-power fit. (c)
Histograms of the TLS and non-TLS loss as a function of aluminum deposition
temperature, showing that the non-TLS loss decreases markedly with increasing
temperature while the TLS loss remains comparable. The inset shows the further
reduction of the TLS loss at \SI{300}{\celsius} across the different etch
profiles.}
\label{fig:loss}
\end{figure}

The two fabrication techniques explored in this work, aluminum microstructural
engineering and silicon dielectric trimming, are shown to consistently improve
the resonator quality factor through loss mitigation. To analyze these losses
further, we model the resonators as an equivalent RLC circuit, illustrated in
Figure~\ref{fig:loss}(a), in which the resistance $R$ represents the combined TLS
and non-TLS dissipation. With the introduced techniques, both contributions are
mitigated, yielding the best resonator performance through \SI{300}{\celsius}
deposition and the Stropic etch (Figure~\ref{fig:loss}(b)), which reaches a
single-photon internal quality factor of $2.3\times10^{6}$ (the inset shows a
representative single-power fit). To identify the microscopic origin of each
contribution, we decompose the total loss into TLS and non-TLS parts,

\begin{equation}
\frac{1}{Q_i} = \frac{1}{Q_{\mathrm{TLS}}} + \frac{1}{Q_{\text{non-TLS}}},
\label{eq:decomp}
\end{equation}

\noindent following the established power-dependent decomposition (see
Supplementary Note~S4) \cite{mcrae2020,altoe2022}. The high-power quality factor
reflects the non-TLS loss, $1/Q_{\text{non-TLS}} = 1/Q_{\mathrm{hp}}$, while the
TLS contribution follows from the difference between the low- and high-power
loss,

\begin{equation}
\frac{1}{Q_{\mathrm{TLS}}} = \frac{1}{Q_{\mathrm{lp}}} - \frac{1}{Q_{\mathrm{hp}}}.
\label{eq:tls}
\end{equation}

Applying this decomposition (Figure~\ref{fig:loss}(c)) reveals a clear separation
of roles: increasing the aluminum deposition temperature, and hence the grain
size, reduces the non-TLS loss while the TLS loss remains essentially constant,
whereas etching the \SI{300}{\celsius} film with different profiles reduces the
TLS loss (inset).

\begin{figure}[!ht]
\centering
\includegraphics[width=0.6\textwidth]{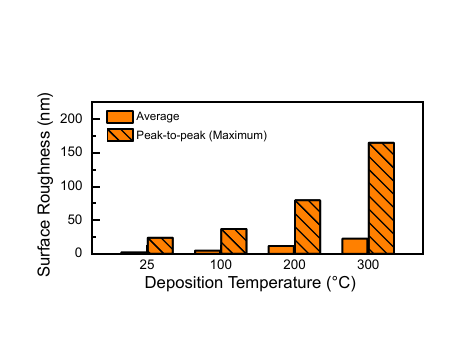}
\caption{AFM measurements of aluminum surface roughness. Surface topography of a
\SI{300}{\celsius} aluminum film. The peak-to-peak roughness is the height
difference between the highest and lowest points, and the average roughness is
the mean deviation from the mean surface height.}
\label{fig:rough}
\end{figure}

This TLS and non-TLS assignment is corroborated by the cross-sectional STEM-EDX
analysis in Figure~\ref{fig:stem}(b). As established there, the grain boundaries
are oxide-free, therefore the loss reduction observed for larger-grain films
originates from the non-TLS channel, consistent with the power-dependent
decomposition in Figure~\ref{fig:loss}(c). These measurements link the metallic
microstructure directly to the non-TLS loss. Larger grains have fewer grain
boundaries, which lowers the non-TLS loss and improves resonator performance.
Notably, this improvement occurs despite an increase in surface roughness. AFM
measurements show that heating the substrate during aluminum deposition
significantly roughens the film surface (Figure~\ref{fig:rough}). We report two
metrics: the peak-to-peak roughness (the height difference between the highest
and lowest points) and the average roughness. The peak-to-peak value rises from
\SI{23}{\nano\meter} at room temperature to roughly \SI{165}{\nano\meter} at
\SI{300}{\celsius}, about 80\% of the \SI{200}{\nano\meter} film thickness, while
the average roughness increases from \SI{2}{\nano\meter} to
\SI{22}{\nano\meter}. Prior work on niobium resonators reports that rougher
surfaces degrade the quality factor \cite{karuppannan2025}. In contrast, we find
that reducing the grain-boundary density improves performance even as the
aluminum surface roughens (Figure~\ref{fig:rough}), indicating that
grain-boundary density, rather than surface roughness, dominates the non-TLS loss
in these films.

The TLS loss, by contrast, is governed by silicon defects at the SA interface and
is therefore reduced by edge etching rather than by the metallic microstructural
changes, consistent with prior work localizing most TLS loss to the SA interface
\cite{altoe2022}. Here we highlight the role of silicon smoothing, at fixed
isotropic SF$_6$ profile, in lowering this loss. High- and low-power quality
factor measurements show that both Tropic profiles reduce the TLS loss, with the
additional sidewall smoothing of the Stropic etch yielding the lowest TLS loss
among all profiles.

As noted in the dielectric-trimming subsection, the etch depth that sets these
profiles also shifts the resonator frequency upward. As a result, our
best-performing devices operate at relatively high frequencies
(6.5--\SI{8.1}{\giga\hertz}), above those of typical quality-factor studies.
High quality factors are harder to maintain at these frequencies because the
density of resonant two-level defects increases with frequency. The
corresponding TLS loss follows

\begin{equation}
\delta_{\mathrm{TLS}}(T) = \delta^{0}_{\mathrm{TLS}}
\tanh\!\left(\frac{\hbar f_{\mathrm{res}}}{2 k_B T}\right).
\label{eq:tlstemp}
\end{equation}

This is consistent with prior aluminum-on-silicon work, where the improvement
from larger grains was not observed at higher frequencies around
\SI{5.9}{\giga\hertz}, and the loss for larger-grain films increased further with
higher frequency \cite{biznarova2024}. The high quality factors maintained here
therefore demonstrate that our loss mitigation remains effective in the regime
where TLS loss is most pronounced.

\section{Conclusion}

Combined, these fabrication techniques provide a route to improve the quality of
both aluminum and silicon, the two materials most widely used in superconducting
circuits. Higher-quality aluminum films directly benefit the ground planes,
capacitor pads, and other circuit elements, and hence the overall coherence of
the quantum circuit. Identifying the silicon bottom and sidewall roughness as a
loss source, in turn, offers a further handle for improving resonator and qubit
performance. The two levers are independent and complementary---microstructural
engineering reduces the non-TLS loss while dielectric trimming reduces the TLS
loss---so they can be used separately or combined. Together, these techniques
push the aluminum-on-silicon platform toward higher coherence and establish
fabrication advancements for superconducting quantum circuits.

\section{Methods}

\subsection*{Device Design and Fabrication}

Each chip contains eight quarter-wavelength ($\lambda/4$) CPW resonators
capacitively coupled to a common transmission line. The resonators have a
center-conductor width $w = \SI{30}{\micro\meter}$ and a gap
$g = \SI{15}{\micro\meter}$, giving a characteristic impedance close to
\SI{50}{\ohm} on the high-resistivity silicon substrate. The designed resonant
frequencies span 5.5--\SI{7}{\giga\hertz}, with coupling quality factors of
$(1.5\text{--}5)\times10^{5}$. The transmission line shares the same
$w\!:\!g = 30\!:\!15$~\si{\micro\meter} geometry. Flux-trapping holes of
\SI{4}{\micro\meter} width and \SI{6}{\micro\meter} pitch surround the
structures.

\subsection*{Fabrication}

Devices were fabricated on high-resistivity ($\rho \geq
\SI{10}{\kilo\ohm\centi\meter}$) silicon substrates. The substrates were cleaned
for \SI{10}{\minute} in piranha solution to remove organic residues, cleaned with
DI water, then followed by a \SI{5}{\minute} dip in buffered oxide etchant (BOE,
7:1) to remove the native oxide at the substrate--metal interface and cleaned
with DI water. The wafer was then loaded into an electron-beam evaporator,
reaching a load-lock pressure of \SI{e-4}{\milli\bar} within \SI{10}{\minute},
and heated to \SI{300}{\celsius} in vacuum to desorb moisture and surface
contaminants before overnight cooling. Once the chamber pressure reached below
\SI{8e-8}{\milli\bar}, a \SI{200}{\nano\meter} aluminum film was deposited at a
rate of \SI{10}{\angstrom\per\second}. To vary the aluminum grain size and
microstructure, the substrate temperature during deposition was varied across
four wafers: 25, 100, 200, and \SI{300}{\celsius}.

The circuits were patterned by photolithography using positive-tone AZ 5214E
resist, and the aluminum was etched in a BCl$_3$/Cl$_2$ reactive-ion-etch (RIE)
process. Following the aluminum etch, the exposed silicon was trimmed under one
of three conditions applied across samples of different grain size, with a fourth
chip left unetched as a baseline. The unetched baseline was only etched with the
BCl$_3$/Cl$_2$ aluminum etch with no deep silicon etch. The anisotropic condition
used a CHF$_3$ RIE to produce vertical smooth sidewalls. The two Tropic
conditions combined an SF$_6$ sidewall etch with a subsequent CHF$_3$ step to
smooth the silicon bottom surface; the sidewall roughness was set by the SF$_6$
chamber pressure, with \SI{60}{\milli\torr} producing the rough-sidewall variant
(Rtropic) and \SI{5}{\milli\torr} the smooth-sidewall variant (Stropic). The
final cleaning of the resist after etching is with acetone spraying for
\SI{15}{\second} followed by NMP at \SI{80}{\celsius} overnight to clean resist
residues. After NMP, the samples are rinsed in IPA and high-power sonication is
used to clean the metal edges. It is important to note that sonication is not
used for the Tropic variants due to the deep undercut etching achieved, since
further sonication causes more metal edge damage.

\subsection*{Cryogenic Measurement Setup}

The resonator chip is wire-bonded to a printed circuit board and enclosed in a
microwave package that is mechanically secured and thermally anchored to the
mixing-chamber stage of a dilution refrigerator by oxygen-free high-conductivity
(OFHC) copper cold fingers. The package is measured at the mixing chamber stage
(\SI{10}{\milli\kelvin}) inside the standard OFHC copper radiation and IR shields
of the refrigerator, without additional magnetic shielding. Measurements are
performed at a base temperature of \SI{10}{\milli\kelvin}.

Cryogenic measurements are carried out with a vector network analyzer (VNA),
which delivers a continuous-wave signal to the device through a semi-rigid
coaxial line. The input line is attenuated by \SI{20}{\decibel} at each of the
\SI{4}{\kelvin}, still, and mixing-chamber stages, with an additional
\SI{43}{\decibel} at room temperature, and \SI{13}{\decibel} for line losses,
giving a total input attenuation of approximately \SI{116}{\decibel}; this places
the resonators in the single-photon regime at the lowest drive powers. At the
mixing chamber, the signal passes through an infrared filter, a circulator, and a
band-pass filter before reaching the device under test. The reflected signal
returns through the circulator, an infrared filter, and an isolator, and is
amplified by a cryogenic high-electron-mobility-transistor (HEMT) amplifier at
the \SI{4}{\kelvin} stage, followed by a room-temperature amplifier, before being
analyzed by the VNA. The output line includes a \SI{1}{\decibel} attenuator at
the \SI{50}{\kelvin} stage, as shown in Supplementary Note~S5.

\subsection*{Atomic Force Microscopy Measurements}

Surface topography of the aluminum conductor films was characterized using a
Bruker Dimension Icon atomic force microscope operated in tapping mode.
Measurements were performed using an ARROW-NCR AFM cantilever at a tapping
frequency of approximately \SI{285}{\kilo\hertz}. Scan areas of
$7.5\times7.5$~\si{\micro\meter\squared}, $10\times10$~\si{\micro\meter\squared},
$15\times15$~\si{\micro\meter\squared}, and
$15\times15$~\si{\micro\meter\squared} were used for Al films deposited at room
temperature, \SI{100}{\celsius}, \SI{200}{\celsius}, and \SI{300}{\celsius},
respectively. Measured AFM topography maps were processed and analyzed using
Gwyddion software \cite{necas2012}. Surface roughness was evaluated from the
maximum height of the profile, $R_t$, defined as the vertical distance between
the highest peak and the deepest valley along a selected profile, following the
profile roughness terminology defined in ISO 21920-2:2021 \cite{iso21920}. For
each AFM image, $R_t$ was calculated row-wise across the topography map, and the
reported roughness value was obtained by averaging $R_t$ over all line profiles.
Grain segmentation was then performed using a watershed-based algorithm
\cite{gwyddionguide}. After segmentation, the characteristic grain size was
evaluated for each identified grain using the maximum bounding size,
$D_{\mathrm{max}}$, and the resulting values were used to obtain the statistical
grain-size distribution.

\section*{Data Availability}

All figures are the original work of the authors and do not include previously
created third-party elements. The data that support the findings of this study
are available from the corresponding author upon request.

\section*{Acknowledgments}

The authors gratefully acknowledge the National Company of Telecommunications and
Information Security (NTIS) for its generous funding and support of this
research, and King Abdullah University of Science and Technology (KAUST) for
additional funding. We thank the KAUST Core Labs team and the nanofabrication
staff for their assistance, and we are especially grateful to Dr.~Menouir Saidani
and Dr.~Syed Kazmi for valuable scientific discussions.

\section*{Funding}

This work was supported by the National Company of Telecommunications and
Information Security (NTIS) and King Abdullah University of Science and
Technology (KAUST).

\section*{Author Contributions}

Conceptualization: M.A., A.S. (Albaraa Shafi), A.H. Fabrication: M.A., U.A.,
A.A. SEM measurements: M.A., A.A., U.A. Cryogenic measurements: I.A.,
A.S.~(Albaraa Shafi). AFM measurements: I.G. TEM measurements: I.G. Data
analysis: M.A., I.A., A.S.~(Albaraa Shafi). Supervision: A.S.~(Atif Shamim).
Writing---original draft and editing: M.A., A.H., A.S.~(Atif Shamim).
Writing---review: all authors.

\section*{Competing Interests}

The authors declare no competing interests.

\end{document}